\documentclass{article}

\usepackage{PRIMEarxiv}

\usepackage[utf8]{inputenc} %
\usepackage[T1]{fontenc}    %
\usepackage{hyperref}       %
\usepackage{url}            %
\usepackage{booktabs}       %
\usepackage{amsfonts}       %
\usepackage{nicefrac}       %
\usepackage{microtype}      %
\usepackage{lipsum}
\usepackage{fancyhdr}       %
\usepackage{graphicx}       %
\usepackage{natbib}
\usepackage{amsmath} 
\title{Signal Enhancement in Distributed Acoustic Sensing Data Using a Guided Unsupervised Deep Learning Network
\thanks{\textit{\underline{Citation}}: 
\textbf{ }} 
}

\author{
  Omar M. Saad, Matteo Ravasi, and Tariq Alkhalifah \\
  Division of Physical Science and Engineering, \\King Abdullah University of Science and Technology (KAUST), Thuwal 23955-6900, Saudi Arabia. \\
}

\begin{document}
\maketitle

\begin{abstract}
Distributed Acoustic Sensing (DAS) is a promising technology introducing a new paradigm in the acquisition of high-resolution seismic data. However, DAS data often show weak signals compared to the background noise, especially in tough installation environments. In this study, we propose a new approach to denoise DAS data that leverages an unsupervised deep learning (DL) model, eliminating the need for labeled training data. The DL model aims to reconstruct the DAS signal while simultaneously attenuating DAS noise. The input DAS data undergo band-pass filtering to eliminate high-frequency content. Subsequently, a continuous wavelet transform (CWT) is performed, and the finest scale is used to guide the DL model in reconstructing the DAS signal. First, we extract 2D patches from both the band-pass filtered data and the CWT scale of the data. Then, these patches are converted using an unrolling mechanism into 1D vectors to form the input of the DL model. The architecture of the proposed DL network is composed of several fully-connected layers. A self-attention layer is further included in each layer to extract the spatial relation between the band-pass filtered data and the CWT scale.  Through an iterative process, the DL model tunes its parameters to suppress DAS noise, with the band-pass filtered data serving as the target for the network. We employ the log cosh as a loss function for the DL model, enhancing its robustness against erratic noise. The denoising performance of the proposed framework is validated using field examples from the San Andreas Fault Observatory at Depth (SAFOD) and Frontier Observatory for Research in Geothermal Energy (FORGE) datasets, where the data are recorded by a fiber-optic cable. Comparative analyses against three benchmark methods reveal the robust denoising performance of the proposed framework.
\end{abstract}

\keywords{Distributed Acoustic Sensing (DAS) \and Denoising \and Unsupervised Deep Learning}

\section{Introduction}
Distributed Acoustic Sensing (DAS) represents a cutting-edge technology with the potential to revolutionize numerous applications, including but not limited to, CO2 monitoring \citep{isaenkov2021automated}, earthquake and microseismic monitoring \citep{lindsey2017fiber,verdon2020microseismic,nayak2021distributed}, earthquake early warning system \citep{farghal2022potential}, and near-surface seismic monitoring \citep{dou2017distributed}. 

DAS systems are composed of a fiber-optic cable and an interrogator unit: the latter sends a laser pulse along the fiber and measures the back-scattered signal originating from inhomogeneities in the fiber \citep{hartog2017introduction}. When the fiber length changes due to, for example, an incoming seismic wave, the back-scattered pulse experiences a phase delay \citep{hartog2017introduction}. The interrogator utilizes a phase modulation technique to transform such phase difference into a relative length change (i.e. strain) over a finite distance, usually referred to as the gauge length \citep{hartog2017introduction,binder2020modeling}. DAS systems excel in producing detailed wavefield information due to their high spatial and temporal resolution compared to conventional geophones \citep{hartog2020distributed,titov2021modeling}. The increased spatial resolution inherent in DAS facilitates its ability to cover long distances, consequently resulting in a diminished cost per sensor for acquisition measurements \citep{binder2020modeling}. Additionally, DAS exhibits robustness in challenging environmental conditions, specifically maintaining its integrity under high-pressure and high-temperature circumstances \citep{booth2020distributed}. 

However, the data obtained from the DAS exhibit a notably low signal-to-noise ratio (SNR) in comparison to conventional geophones \citep{hull2017case}. The noise within the DAS data comes from various sources, including cultural, optical, environmental, and coupling effects. Cultural noise primarily stems from human activities, such as airplanes, footsteps, electrical power lines, and vehicles. Environmental noise contains several factors like winds, thunders, and earth tides. Optical noise, contributing to both horizontal and vertical noise in the DAS data, can arise from malfunctions in the interrogator during the laser emission process. Furthermore, in wellbore applications, imperfect coupling of the fiber-optic cable to the borehole wall can introduce additional noise in certain instances. Therefore, employing a robust denoising framework is crucial for effectively mitigating these complex noise components.

Over the past few decades, several traditional methods have been proposed to attenuate random and coherent seismic noise, e.g., predictive filtering \citep{abma1995lateral}, and the f-x deconvolution method \citep{canales1984random}. Additionally, various methods involve transforming seismic data into alternative domains, such as Fourier transform \citep{fourier1986},  wavelet transform \citep{anvari2017seismic}, and curvelet transform \citep{candes20061}, aiming to suppress undesired components while preserving the signal by applying a threshold function. 

Some of these traditional methods have been utilized to denoise DAS data. \citep{lellouch2021low} proposed to use a median filter and low-pass filter to enhance the SNR of the Frontier Observatory for Research in Geothermal Energy (FORGE) DAS dataset. \citep{binder2020modeling} proposed to utilize the dip filter to attenuate the vertical and horizontal DAS noise. \citep{yu2016walkaway} used the sparse transformation to suppress the coupling noise generated by the instrument. However, these traditional denoising methods suffer from signal leakage and are not well-suited for all the complex noise found in DAS data.

Several supervised deep-learning techniques have been proposed to enhance the SNR of DAS data. For instance, an adversarial denoising network has been proposed to retrieve the weak DAS signal while attenuating the DAS noise \citep{dong2020denoising}. Another approach, proposed by \citep{cheng2023multiscale}, involved using a multiscale recurrent self-attention network to attenuate background noise in seismic data acquired for a vertical seismic profile experiment. \citep{wu2022multi} introduced a multiscale fusion attention network, trained on a smaller dataset, to enhance the SNR of DAS data. \citep{li2023attention} proposed using an attention convolutional neural network (CNN) to address complex DAS noise. Additionally, \citep{yang2023denoising} used a dense network with residual connections to effectively attenuate DAS noise, while \citep{zhao2022coupled} employed CNN to suppress the coupled noise in DAS data.

Although these methods have shown promise in reducing noise, they require training involving synthetic data with a sufficiently large number of labels. To address this limitation, some models have been proposed using self-supervised and unsupervised approaches to suppress DAS noise \citep{yang2023denoising1, zhao2023self}. Nonetheless, it is imperative to recognize that the denoising task invariably involves a trade-off between reconstructing the signal and attenuating the noise.

In this work, we propose an unsupervised framework for attenuating the DAS noise, where no labels are required. This framework uses continuous wavelet transform (CWT) to obtain diverse scales of DAS data, guiding a deep learning network in reconstructing the DAS signal while reducing noise. Through an unsupervised approach, the deep learning model iteratively denoises the band-pass filtered data, adapting its parameters for optimal performance. Finally, a dip filter in the f-k domain efficiently eliminates residual horizontal DAS noise that could result from the patching. We will demonstrate the effectiveness of this proposed method on multiple field data including data used for Earthquake monitoring.

The paper is organized as follows. Following these introductory statements, we share the theoretical development, which includes the 2D CWT transform and its integration into a novel neural network architecture. Results of the application of this method on multiple field DAS data will be shared next. In the discussions, we will delve into the features of the used network architecture, including the role of its components, the training set, computational complexity, and try to interpret the model's behavior. We also share the limitations of the proposed method and share our concluding remarks.

\section{Methodology}\label{sec:Methodology}
The proposed framework adopts an unsupervised approach, eliminating the necessity for labeled data. Firstly, the 2D DAS data undergo band-pass filtering to eliminate the high-frequency content. Subsequently, a median filter with a kernel size of 3 is applied to the band-pass filtered data to facilitate signal smoothing. Following this, the scales for the 2D CWT are determined from the smoothed data. 

We employ a patching technique where the 2D data are partitioned into overlapping windows, each sized $S1 \times S2$, with a shift of $l1 \times l2$ samples between neighboring windows. Each patch is converted to a 1D vector. Consequently, the input to the deep learning (DL) network is the concatenation of the patches extracted from the finest scale of the CWT and the band-pass filtered data as shown in Figure \ref{fig1}. In the reconstruction phase, where denoised patches from the proposed DL network are merged back into the 2D data, an averaging process is applied to the overlapped samples of neighboring windows.

\subsection{2D Continuous Wavelet Transform}

The DL network reconstructs the DAS signal with guidance from the finest scale of the 2D CWT. The 2D CWT, an extension of the 1D CWT, achieves enhanced efficiency by performing the transformation in the Fourier space \citep{antoine2008two,wang2010two}. This involves using 2D fast Fourier transforms on the signal. Afterward, the obtained signal is multiplied by the Fourier transform of the wavelet at the chosen scale, which constitutes a convolution in the time domain. The 2D CWT in the Fourier space ($W_f$) can be obtained as follows \citep{antoine2008two,wang2010two}:
\begin{equation}\label{eq:eq1}
W_f(\textbf{b},s) = \int_{R^2} \frac {1}{s} \hat{y}(\textbf{w}) \hat{\bar{\psi}} \left[\textbf{R}_{\theta}^{-1} \left(\frac{\textbf{w}}{s}\right) \right] e^{i\textbf{b}\textbf{w}} d\textbf{w}
\end{equation}
where $s$ represents the scale parameter, $\textbf{R}_ \theta$ is the rotation matrix with an angle of $\theta$, and $\textbf{b}$ is the position vector \citep{antoine2008two,wang2010two}. The variable $y$ stands for the 2D input data, with ($~\hat{}~$) representing the Fourier transform, and ($~\bar{}~$) indicating the complex conjugate of the function. The 2D wavelet mother function in the frequency domain can be obtained as follows \citep{wang2010two}:
\begin{equation}\label{eq:eq2}
\hat{\psi}_{s,\theta}(\textbf{w}) = s \hat{\psi} \left[s\textbf{R}_{\theta}^{-1} (\textbf{w}) \right].
\end{equation}
Given that the 2D Mexican hat wavelet is derived from the Laplacian of the 2D Gaussian function \citep{marr1980theory,wang2010two}, we utilize it to serve as the mother wavelet to smooth the DAS data. This utilization aims to guide the proposed DL network in the extraction of the DAS signal. 

The Mexican hat wavelet is employed with 19 scales, as shown in Figure \ref{fig2}. The frequency range exhibits a gradual decrease from a larger scale (the $1^{st}$ scale) to the smallest one (the $19^{th}$ scale). Notably, the coefficients at the latest scale exhibit a smoother shape compared to the initial scales. In particular, at the $19^{st}$ scale, the DAS signal becomes prominently visible in contrast to the $1^{st}$ scale. Therefore, we choose the coefficients from the $19^{th}$ scale to guide the proposed DL network in retrieving the DAS signal.

\subsection{The Proposed Deep Learning Network}
\begin{figure}[h] 
\centering
\includegraphics[scale=0.5]{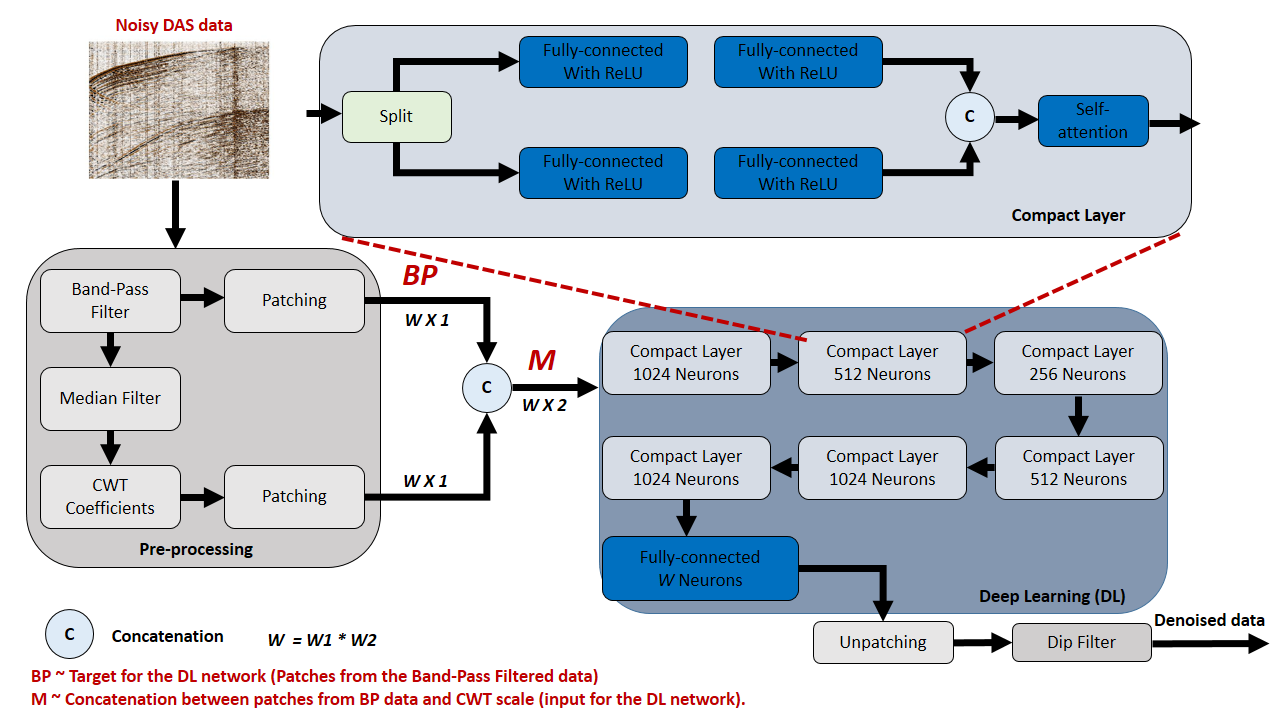} 
\caption{The proposed framework for DAS denoising.  W1 and W2 are the width and height of the patch size.} 
\label{fig1} 
\end{figure}

\begin{figure}[h] 
\centering
\includegraphics[scale=0.5]{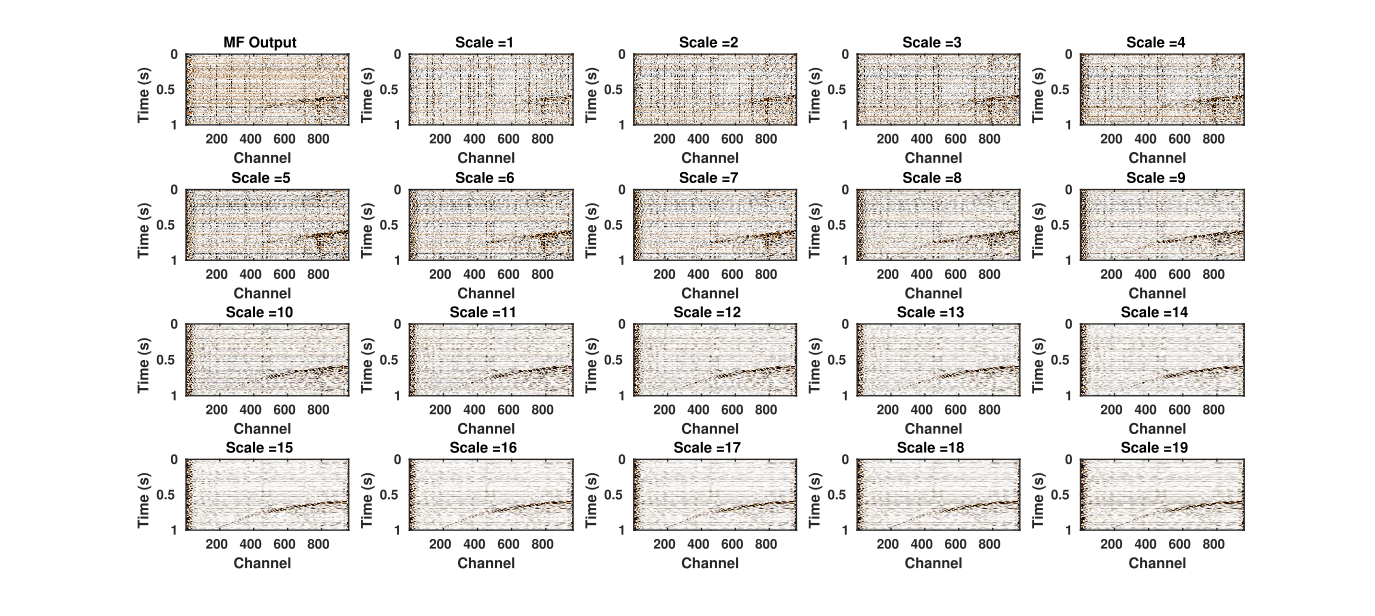} 
\caption{The scale coefficients of the Mexican hat wavelet. The input to the CWT is the output of the median filter, and we use the latest scale coefficients (the $19^{th}$ scale) as input for the DL model.} 
\label{fig2} 
\end{figure}

The inputs for the proposed DL network comprise 1D patches obtained from the band-pass filter and the finest scale of the CWT. The proposed DL network denoises the DAS data in an unsupervised manner. A notable strength of this framework is its fixed patch size, allowing adaptability across diverse data.  The architecture of the DL network is paramount for achieving robust denoising performance. Since the input of the DL network is 1D, we use several fully connected layers as shown in Figure \ref{fig1}. The proposed DL network consists of three encoder and decoder layers, each featuring a compact layer. The input to the compact layer includes two channels—one for the band-pass filtered patches, $z$, and another for the scale coefficients patches, $s$. Initially, the compact layer uses a split block to separate the two channels into two branches as shown in Figure \ref{fig1}. Each branch of the compact layer consists of two fully connected layers, and the output of each branch is determined as follows:
\begin{equation}\label{eq:eq3}
\begin{split}
O_p(z) = \sigma (W_{p2} \times \sigma (W_{p1} \times z + b_{p1}) + b_{p2}),\\
O_s(s) = \sigma (W_{s2} \times \sigma (W_{s1} \times s + b_{s1}) + b_{s2}),\\
\end{split}
\end{equation}
where $\sigma$ denotes the rectified linear unit (ReLU), $W_{p1}$ and $W_{p2}$ are the weight matrices of the two fully connected layers on the band-pass filter-branch. Correspondingly, $W_{s1}$ and $W_{s2}$ denote the weight matrices for the two fully connected layers in the scale branch. The biases for the first fully connected layers in the two branches are represented by $b_{p1}$ and $b_{s1}$, while $b_{p2}$ and $b_{s2}$ denote the biases for the second fully connected layers in both branches. The outputs of the two branches, denoted as $O_p(z)$ and $O_s(z)$, are concatenated to form two separate channels, subsequently passing through a self-attention network. The four fully connected layers in the same compact layer have the same number of neurons. We set the patch size to $48 \times 48$ with shift samples of $8 \times 8$. The number of neurons for the three encoder layers is 1024, 512, and 256, from shallow to deep, while the three decoder layers employ 512, 1024, and 1024 neurons, respectively. 

The attention network aims to extract the spatial relationships among features derived from both branches. The self-attention network achieves this by assigning higher attention weights to the significant features within its input. Since the scale branch provides a smoother representation of the DAS data, the attention network extracts features that are particularly relevant to the DAS signal, emphasizing the importance of these significant features. The attention network comprises three layers, namely, 1D global average, and two fully connected layers incorporating ReLU and sigmoid activation functions, in order \citep{hu2018squeeze}. The 1D global average is performed on the spatial dimension, enabling the attention network to capture the interdependence and mutual relations between the two branches. The final output of the attention network is obtained by multiplying the input of the attention network ($r$) with the output from its second fully connected layer ($V$) which can be represented as follows \citep{hu2018squeeze}:
\begin{equation}\label{eq:eq4}
A(r) = r \times V.
\end{equation}
The output of the second fully connected layer, denoted as $V$, can be obtained using:
\begin{equation}\label{eq:eq5}
V = \delta (W_{a2}~ \times ~  \sigma(W_{a1} \times U)~),
\end{equation}
where $\delta$ is the sigmoid function, and $U$ denotes the output of the 1D average pooling layer, which is obtained as follows:
\begin{equation}\label{eq:eq6}
U = \frac {1}{K}\sum_{i=1}^{K} r(i),
\end{equation}
where $K$ is the input vector ($r$) length.

Following the third decoder layer, we include an extra fully connected layer with the same input size. This layer employs a linear activation function and serves as the output layer. For optimizing the parameters of the proposed network, we employ the Adam optimizer \citep{kingma2014adam} with an initial learning rate of 1e-3, which is reduced to 5e-4 after the $50^{th}$ epoch. The optimization process spans 100 epochs, and it is terminated when the loss fails to decrease for five consecutive epochs. We adopt the log cosh \citep{saleh2022statistical} as a loss function which is defined as follows:
\begin{equation}\label{eq:eq7}
loss = log( \frac{e^{N(M;\phi) - BP} + e^{- (N(M;\phi) - BP)} } {2} ),
\end{equation}
where $N$ and $\phi$ denote the proposed DL network and its parameters, respectively. Meanwhile, $M$ and $BP$ denote the input patches of the DL network and the band-pass filtered patches, respectively. The expression $N(M;\phi) - BP$ represents the error between the predicted and the target patches. While the log cosh operates similarly to a mean square error, it exhibits resilience against occasional highly inaccurate predictions (outliers) \citep{chen2018log}. Therefore, the log cosh loss function is beneficial for the network as it helps diminish the impact of strong amplitude changes, particularly in the presence of erratic vertical noise.

Following the reconstruction of the 2D DAS data using the unpatching technique, the f-k dip filter is employed to eliminate any residual horizontal noise. The f-k dip filter utilizes a single parameter, denoted as $c$, which represents the width of the triangle mask filter in percentage \citep{chen2023denoising}. The permissible range for $c$ spans from 0 to half the width of the triangle mask (0.5). A selected small value for $c$, specifically 0.02, proves effective in successfully removing the residual horizontal noise \citep{chen2023denoising}.

\section{Examples}\label{sec:Examples}
To evaluate the performance of the proposed framework, four field examples are used corresponding to data from FORGE and the San Andreas Fault Observatory at Depth (SAFOD) sites. A subsequent comparison of the results is performed with benchmark methods for a comprehensive evaluation.

\subsection{FORGE data}
For the FORGE data, a fiber cable was deployed for a 10.5-day monitoring period in May 2019. The fiber optic has a channel spacing of 1 m and 10 m gauge length. The sampling interval of the FORGE data is 0.0005 s. Similar to \citep{chen2023denoising}, we leverage the earthquake catalog provided by \citep{lellouch2021low} for extracting a relevant portion of the recorded DAS data. This process results in the extraction of 960 channels, each comprising 2000 samples, starting with 50 samples before the P-wave arrival time of an event in the catalog.

\subsection{First Field Example}
\begin{figure}[h] 
\centering
\includegraphics[scale=0.3]{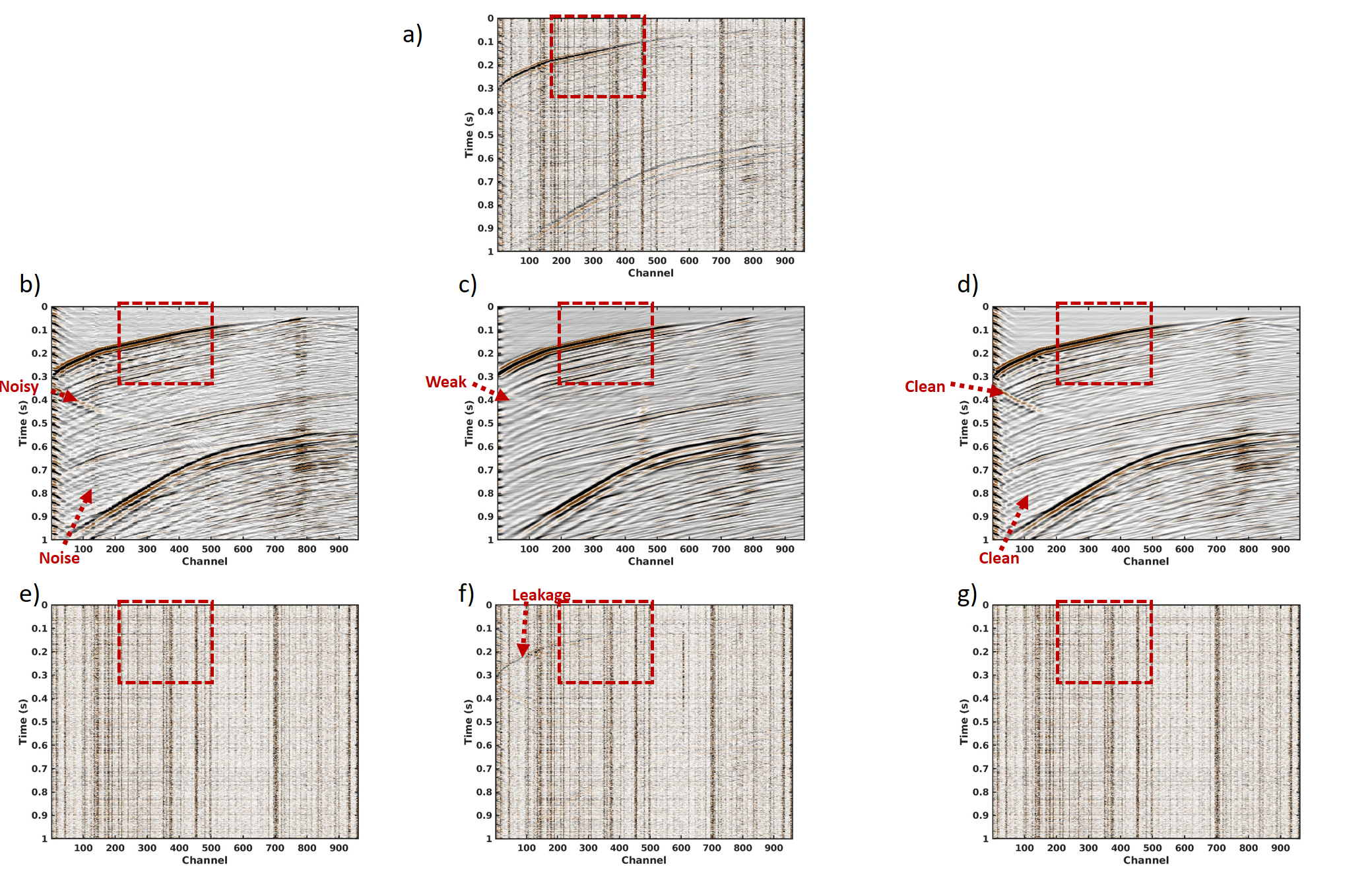} 
\caption{Denoising results for the first field data example from the FORGE site. (a) Noisy DAS data. The denoised data corresponing to the (b) SOMF method, (c) FCDensenet, and (d) the proposed framework. The removed noise section corresponding to the (e) SOMF method, (f) FCDensenet, and (g) the proposed framework.} 
\label{fig3} 
\end{figure}

Figure \ref{fig3} (a) shows a 1 second time window that contains major events attributed to an earthquake, stored in SEGY format under the file name "FORGE\_78-32\_iDASv3-P11\_UTC190502024438.sgy" \citep{lellouch2021low}. This field DAS data correspond to an earthquake with a local magnitude of 0.065 \citep{lellouch2021low}. The extracted data, sized $2000 \times 960$, are band-pass filtered between 1e-3 and 250 Hz. The smallest CWT scale is determined as the input for the DL network, with the band-pass filtered data serving as the target. Figure \ref{fig3}(d) shows the denoised data obtained by the proposed framework, while the removed noise section is shown in Figure \ref{fig3} (g). 

For comparative analysis, two unsupervised methods, namely FCDensenet \citep{yang2023denoising1} and SOMF \citep{chen2023denoising}, are employed. FCDensenet utilizes an unsupervised deep learning approach, while SOMF relies on a structure-oriented median filter. Figures \ref{fig3}(b) and (c) show the denoised data corresponding to the SOMF and FCDensenet, respectively, and the removed noise sections by the two methods are shown in Figures \ref{fig3}(e) and (f). 

\begin{figure}[h] 
\centering
\includegraphics[scale=0.3]{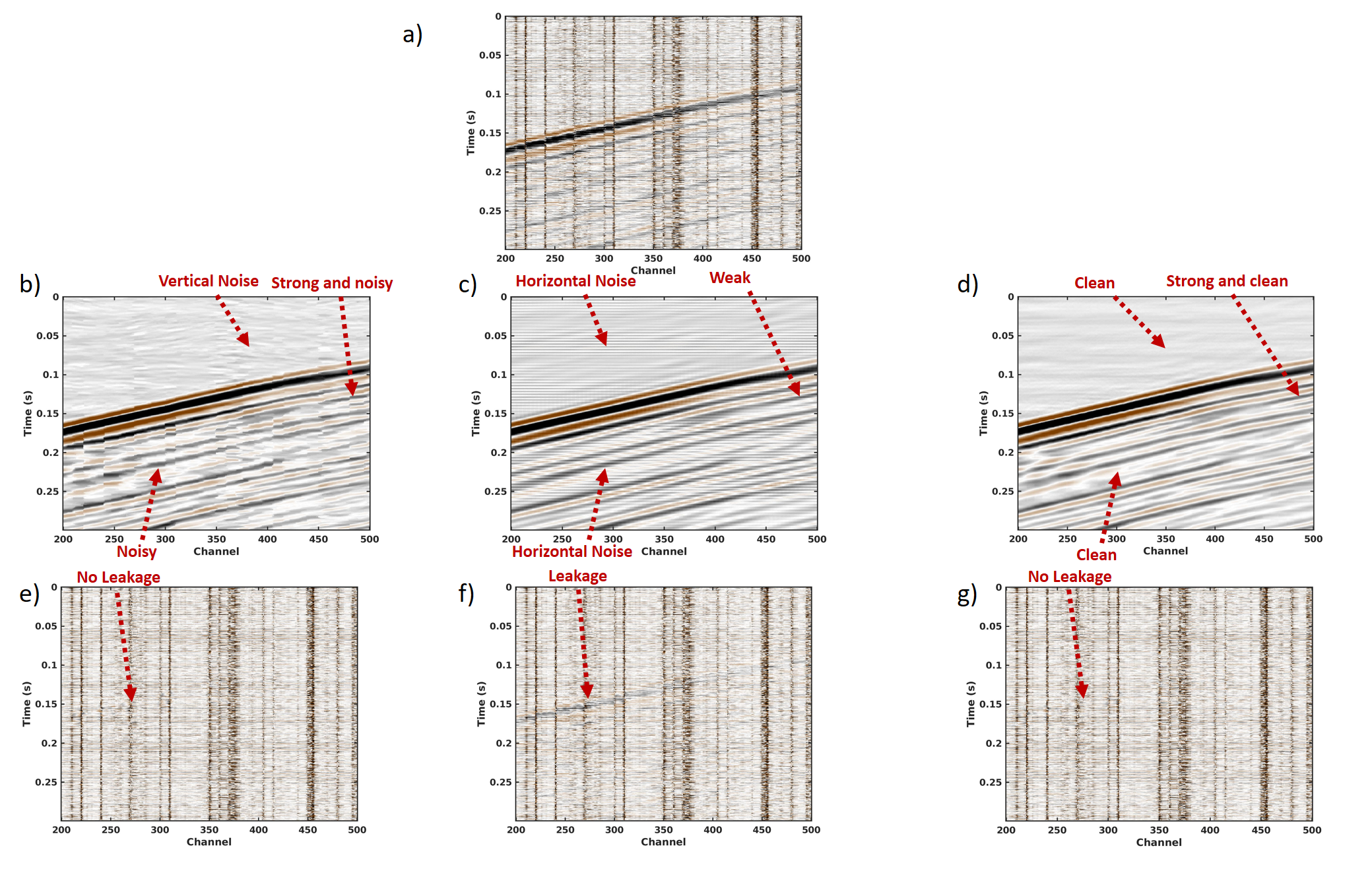} 
\caption{Zoomed sections from the red rectangles in Figure \ref{fig3}. (a) Noisy DAS data. The denoised data corresponding to the (b) SOMF method, (c) FCDensenet, and (d) the proposed framework. The removed noise section corresponding to the (e) SOMF method, (f) FCDensenet, and (g) the proposed framework.} 
\label{fig4} 
\end{figure}

Figure \ref{fig4} shows zoomed-in sections corresponding to the highlighted regions outlined by red rectangles in Figure \ref{fig3}. According to Figures \ref{fig3} and \ref{fig4}, the FCDensenet attenuates most of the noise. However, it suffers from signal leakage  (see Figures \ref{fig3}(f) and \ref{fig4}(f)), and it fails to remove the horizontal noise as shown in Figures \ref{fig3}(c) and \ref{fig4}(c). In contrast, the SOMF method reconstructs the signal with minimal leakage but retains some residual noise, as observable in Figures \ref{fig3}(b) and \ref{fig4}(b).

The proposed algorithm exhibits robust performance when compared to the benchmark methods, particularly in terms of minimizing signal leakage and attenuating noise. Figures \ref{fig4}(d) and (g) highlight the capability of the proposed framework to reduce DAS noise without inducing signal leakage. In comparison to the benchmark methods, the proposed algorithm yields the cleanest denoised data.

\begin{figure}[h] 
\centering
\includegraphics[scale=0.3]{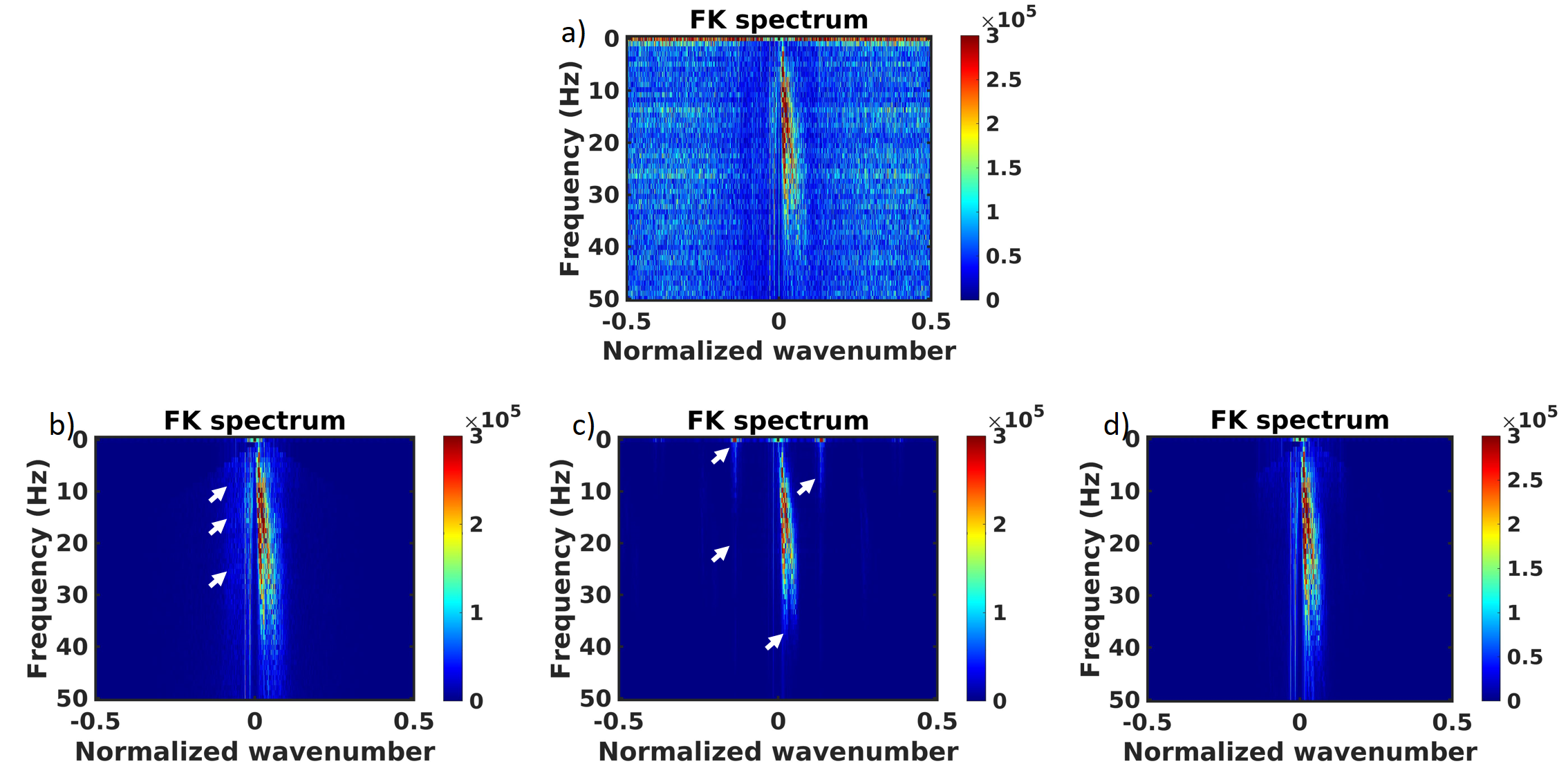} 
\caption{The f-k spectrum for the (a) noisy DAS data, (b) SMOF method, (c) FCDensenet, and (d) the proposed framework. This f-k spectrum corresponding to the field data in Figure \ref{fig3}} 
\label{fig5} 
\end{figure}

Furthermore, we plot the f-k spectrum for the noisy data and the denoised data from the three methods as shown in Figure \ref{fig5}. In the f-k spectrum of the SOMF and the FCDensenet, some residual vertical and horizontal noise is noticeable, marked by arrows. However, the f-k spectrum corresponding to the proposed framework does not show residual noise. Additionally, we observe that the f-k spectrum from FCDensenet loses certain frequency contents related to the DAS signal. From these results, we can conclude that the proposed framework outperforms the benchmark methods in attenuating DAS noise and preserving the DAS signal.

\subsection{Second Field Example}

\begin{figure}[h] 
\centering
\includegraphics[scale=0.35]{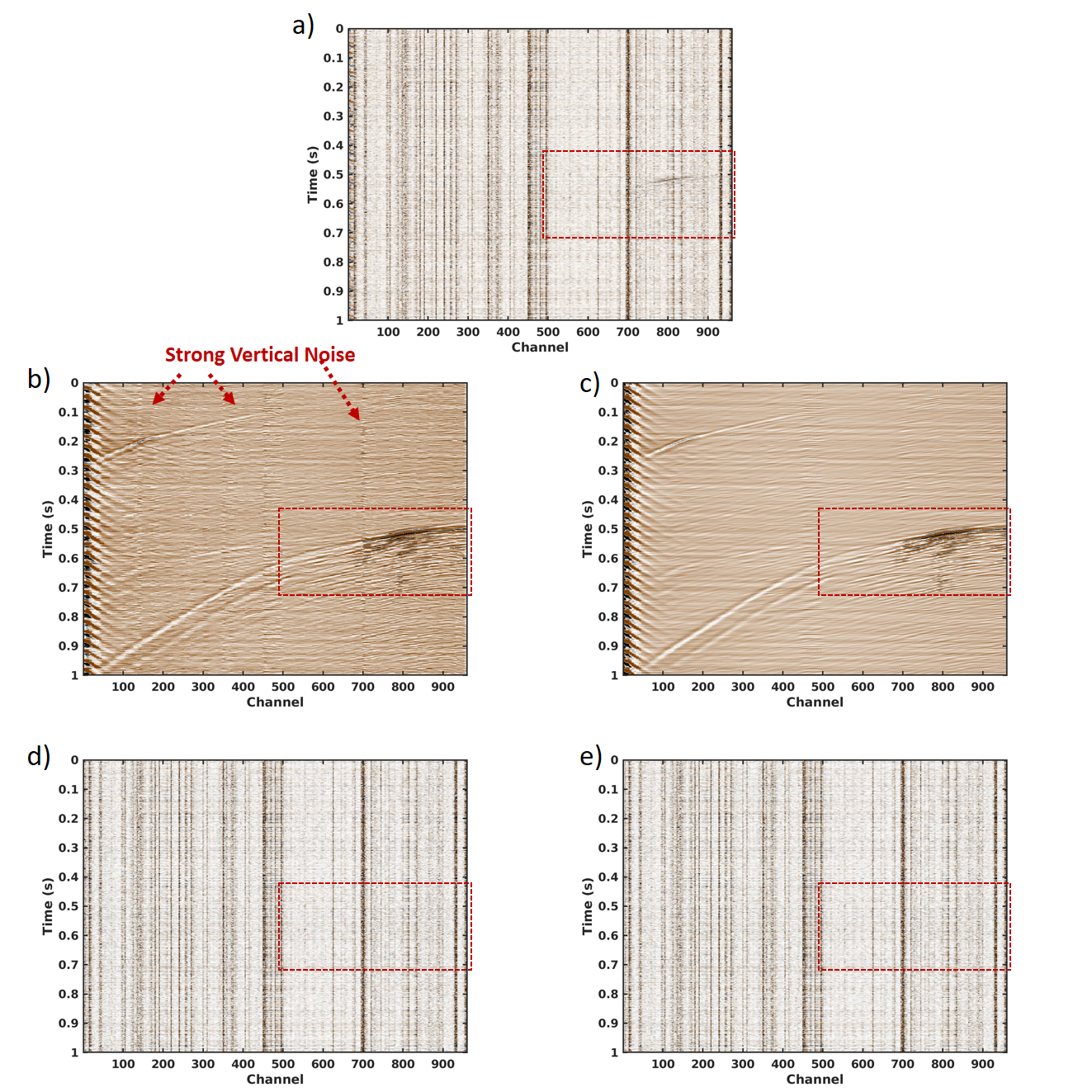} 
\caption{Denoising results for the second field example data from the FORGE site. (a) Noisy DAS data. The denoised data corresponding to the (b) SOMF method and (c) the proposed framework. The removed noise section corresponding to the (d) SOMF method and (e) the proposed framework.} 
\label{fig6} 
\end{figure}

The second field example of DAS data, captured during an earthquake event with a local magnitude of -0.087 \citep{lellouch2021low}, is shown in Figure \ref{fig6} with the file name "FORGE\_78-32\_iDASv3-P11\_UTC190428175708.sgy". Denoised data from the SOMF method and the proposed framework are shown in Figures \ref{fig6}(b) and (c), respectively. The corresponding removed noise sections are presented in Figures \ref{fig6}(d) and (e).

\begin{figure}[h] 
\centering
\includegraphics[scale=0.35]{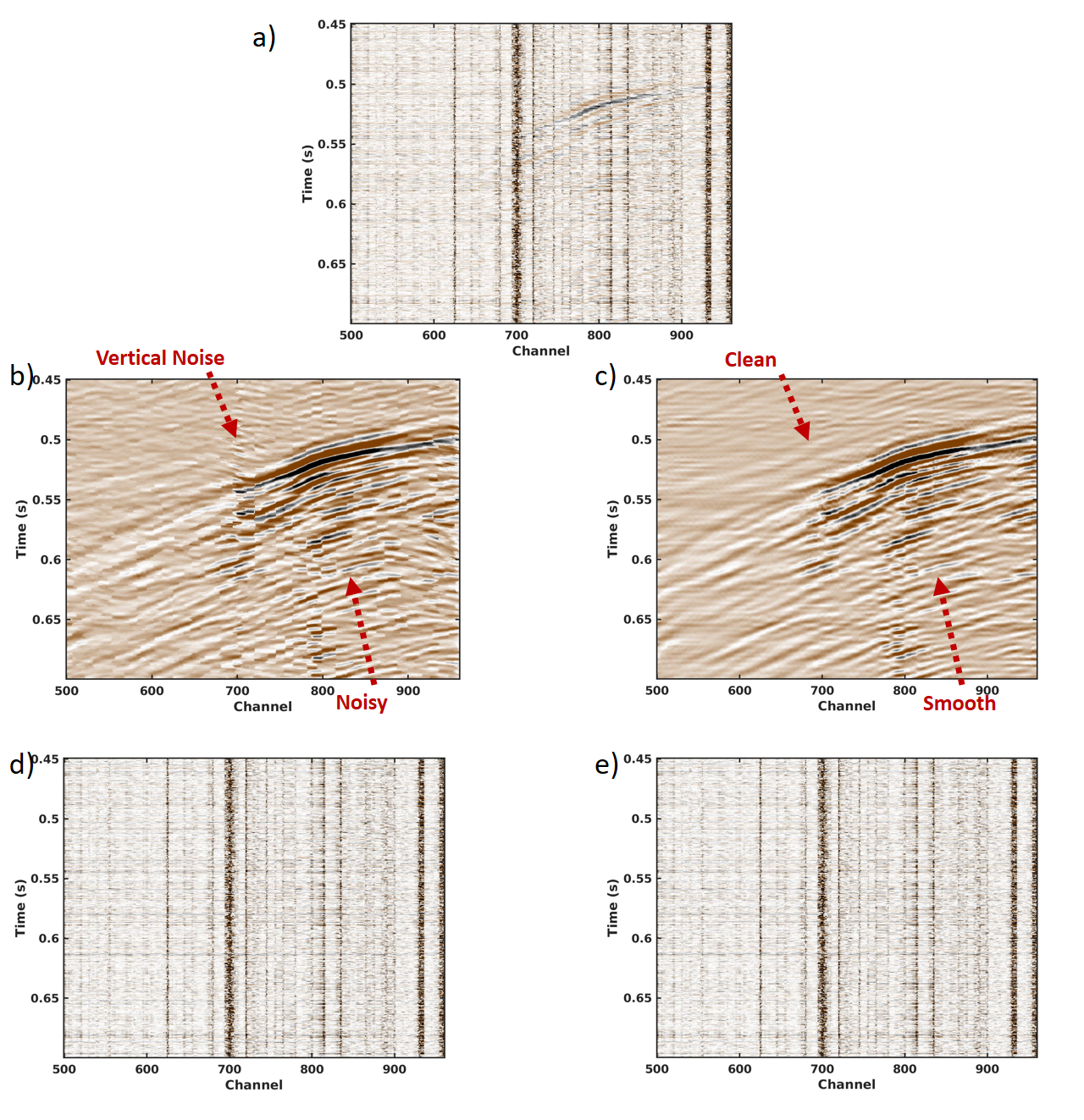} 
\caption{Zoomed sections from the red rectangles in Figure \ref{fig6}. (a) Noisy DAS data. The denoised data corresponding to the (b) SOMF method, and (c) the proposed framework. The removed noise section corresponding to the (d) SOMF method and (e) the proposed framework.} 
\label{fig7} 
\end{figure}

Both methods effectively retrieve the DAS signal with minimal signal leakages as shown in Figures \ref{fig6}(d) and (e). However, the SOMF method retains some residual noise, particularly in the vertical direction, as observed in Figures \ref{fig6}(b) and \ref{fig7}(b). In contrast, the denoised data from the proposed framework is clean and smooth as shown in Figures \ref{fig6}(c) and \ref{fig7}(c). 

\begin{figure}[h] 
\centering
\includegraphics[scale=0.3]{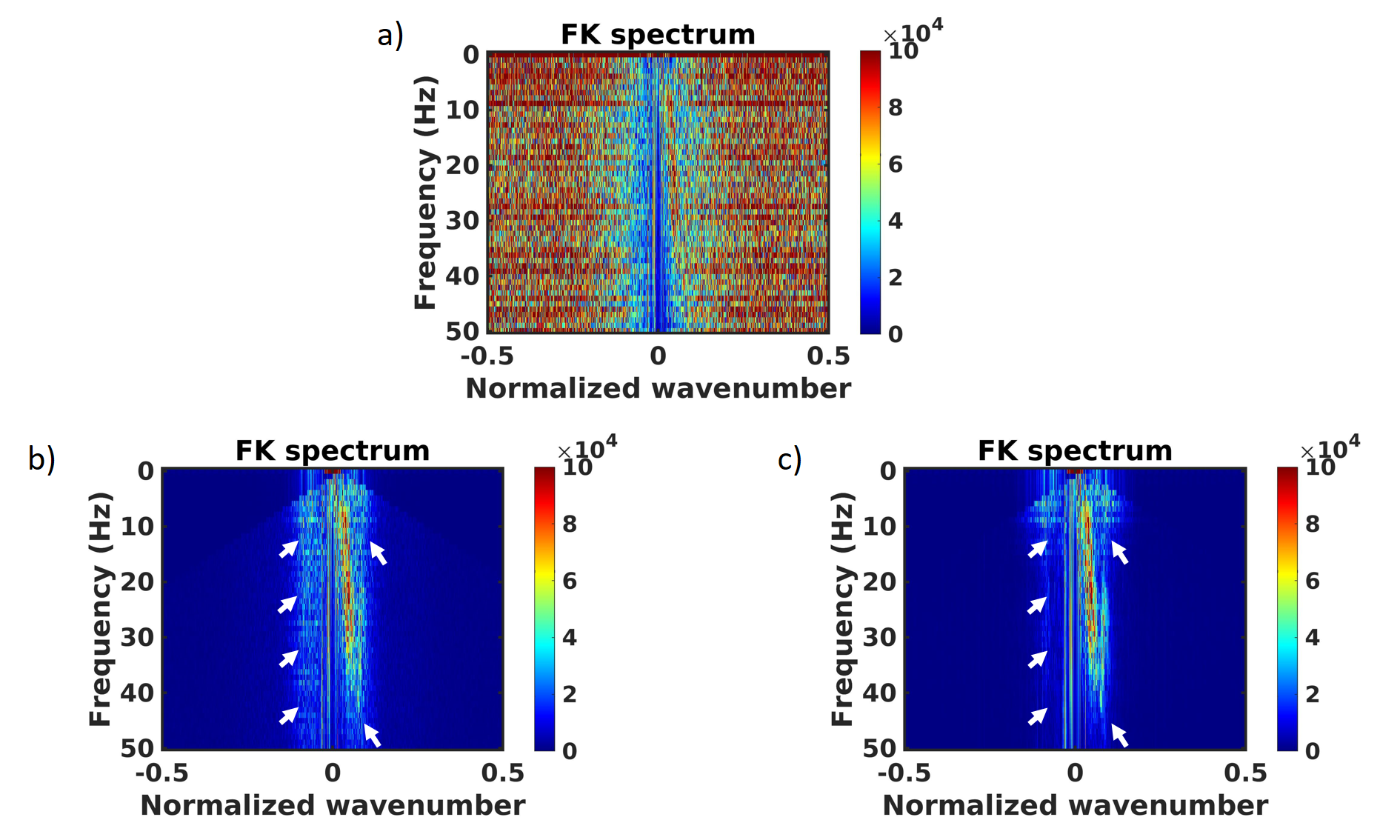} 
\caption{The f-k spectrum for the (a) noisy DAS data, (b) SMOF method, and (c) the proposed framework. These f-k spectra correspond to the field data in Figure \ref{fig6}.} 
\label{fig8} 
\end{figure}

Furthermore, the f-k spectrum of the SOMF retains some frequency contents related to the DAS noise, highlighted by arrows in Figure \ref{fig8}(b), whereas the f-k spectrum corresponding to the proposed framework shows that the proposed method effectively attenuates those noise frequencies, as demonstrated in Figure \ref{fig8}(c). These results indicate that the proposed framework excels in attenuating DAS noise with minimal signal leakages compared to the SOMF method.

Moreover, attempts to denoise this field data using the FCDensenet model were unsuccessful. The FCDensenet model did not converge, even with some tuning to the original architecture, particularly in cases with an extremely low SNR and complex DAS noise, which is the case here.

\subsection{SAFOD data}
In the next subsections, we evaluate the effectiveness of the proposed framework using SAFOD DAS data associated with earthquakes \citep{lellouch2021low}. Two field examples from the SAFOD data are tested using the proposed framework, each sized of $800 \times 14,999$ with a 1-meter channel interval and a 250 Hz sampling rate. We apply the proposed framework to the noisy SAFOD data to suppress complex DAS noise and retrieve the DAS signal. Additionally, we use the moving rank-reduction (MRR) filter \citep{chen2023enhancing} as a benchmark method for result comparison.

\subsection{Third Field Example}
\begin{figure}[h] 
\centering
\includegraphics[scale=0.5]{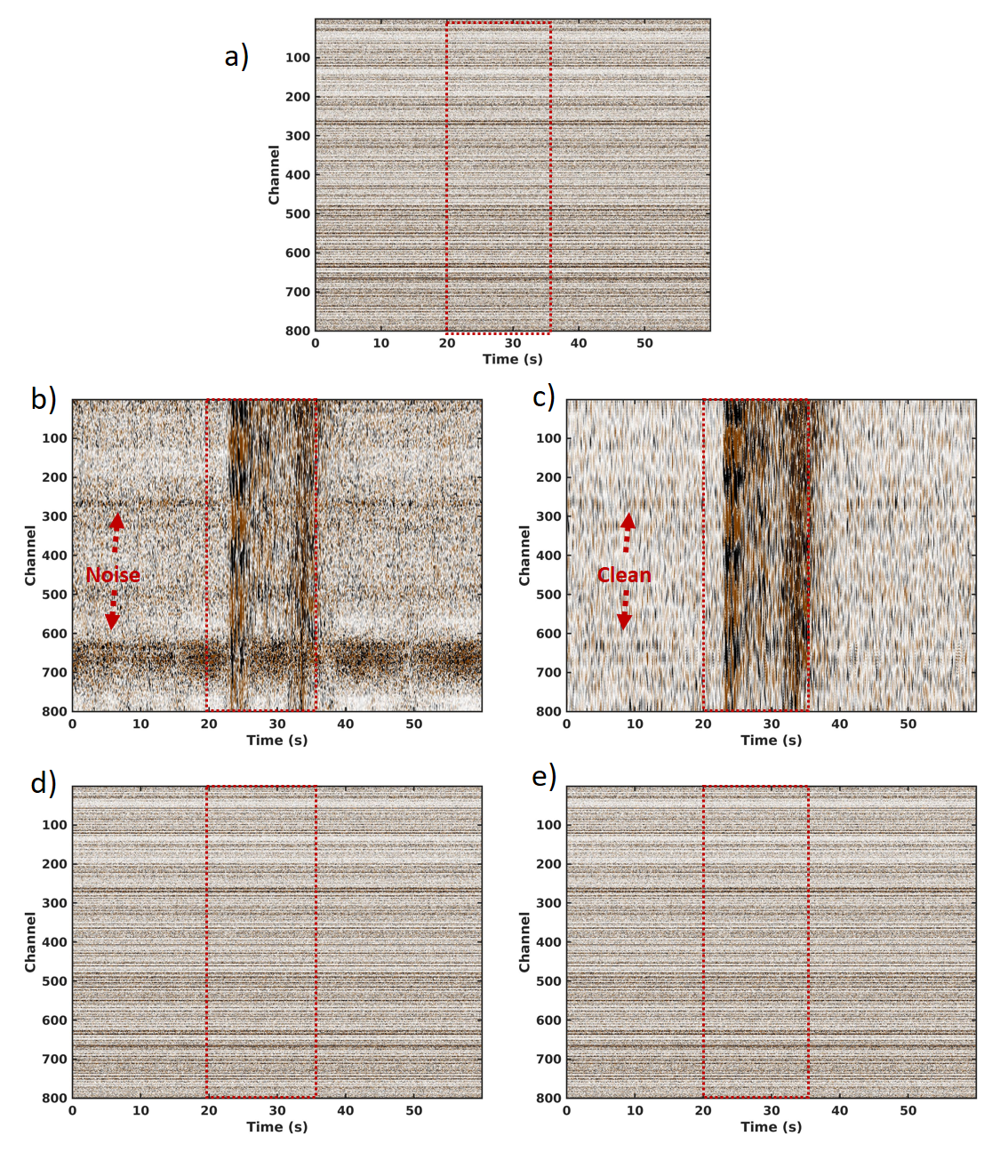} 
\caption{Denoising results for the third field example data from the SAFOD site. (a) Noisy DAS data. The denoised data correspond to the (b) MRR method and (c) the proposed framework. The removed noise sections correspond to the (d) MRR method and (e) the proposed framework.} 
\label{fig9} 
\end{figure}

The third field example pertains to SAFOD DAS data associated with an earthquake with a local magnitude of 2.17 as shown in Figure \ref{fig9}(a) and named '2017-06-23T22:03:22.380000Z\_mag2.17' \citep{lellouch2021low}. Notably, the DAS signal is barely noticeable due to the complex nature of the DAS noise. However, both the MRR method and the proposed framework successfully rendered the DAS signal visible as shown in Figures \ref{fig9}(b) and (c).  Both methods exhibit minimal signal leakage, which can be noticed in the removed noise sections corresponding to both methods in Figures \ref{fig9}(d) and (e). 

\begin{figure}[h] 
\centering
\includegraphics[scale=0.5]{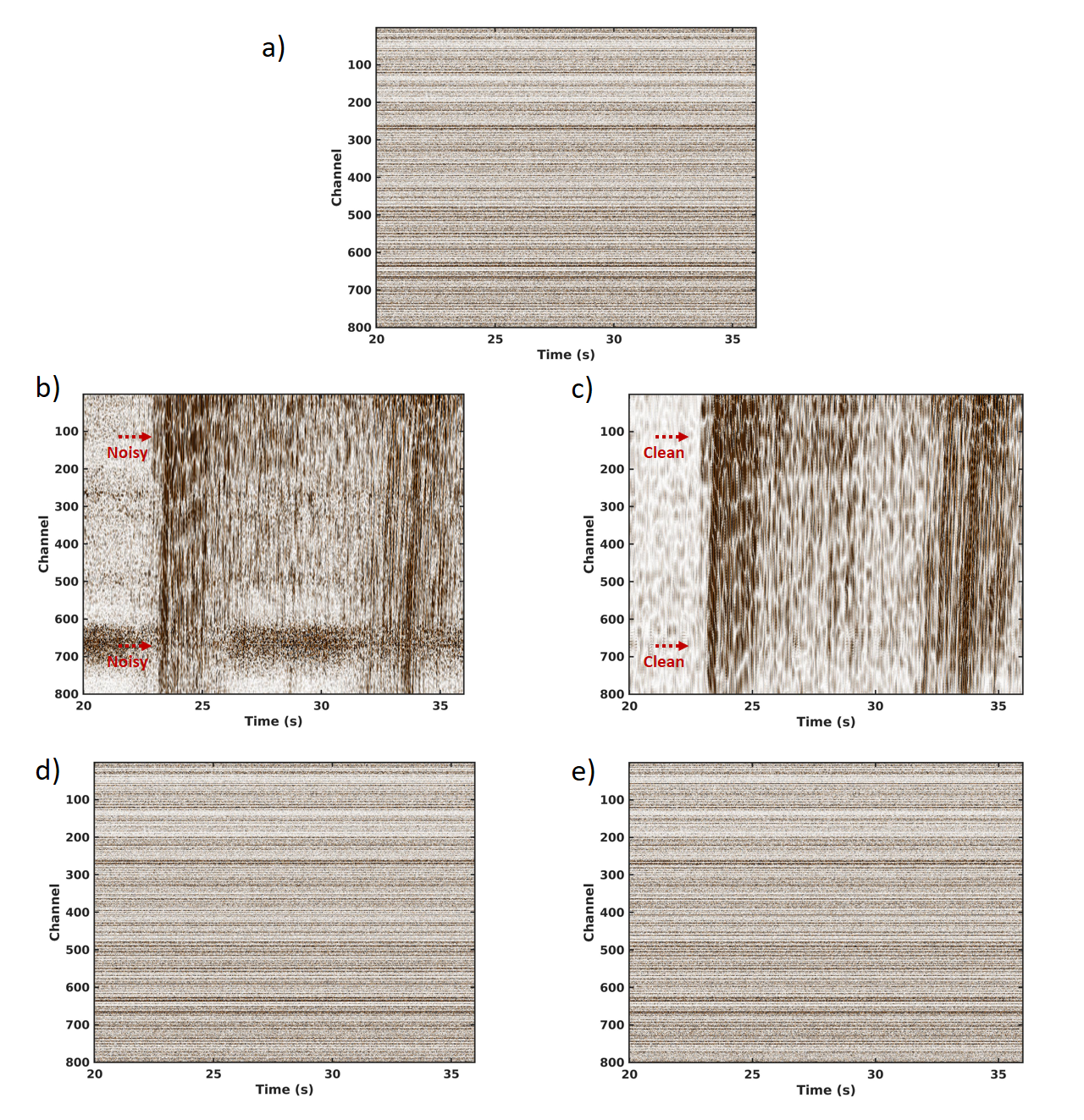} 
\caption{Zoomed sections from the red rectangles in Figure \ref{fig9}. (a) Noisy DAS data. The denoised data correspond to the (b) MRR method, and (c) the proposed framework. The removed noise sections correspond to the (d) MRR method and (e) the proposed framework.} 
\label{fig10} 
\end{figure}

\begin{figure}[h] 
\centering
\includegraphics[scale=0.3]{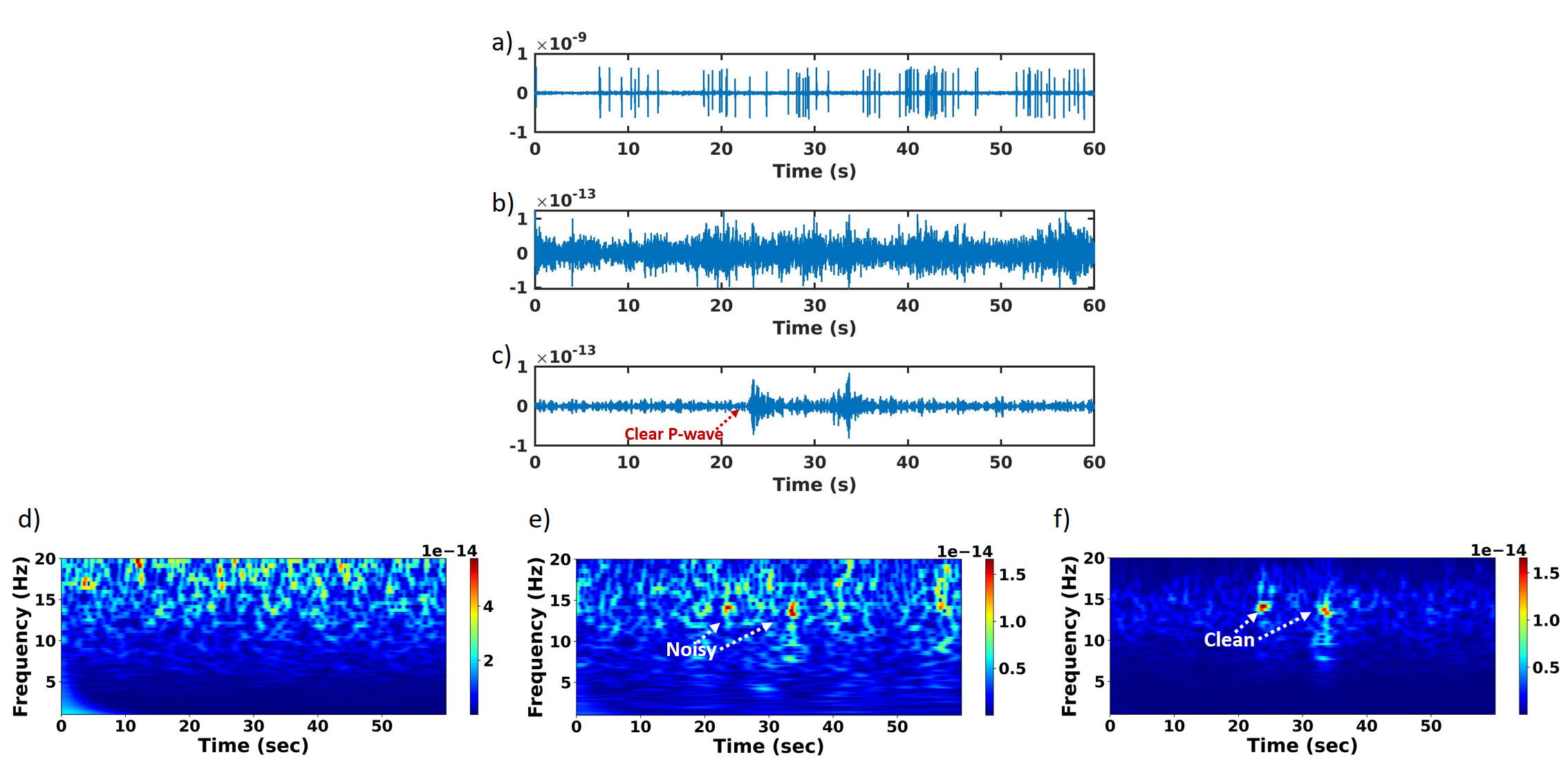} 
\caption{Single-trace comparison. (a) The 70 $^{th}$ noisy trace in Figure \ref{fig9}. The denoised traces correspond to (b) the MRR method and (c) the proposed framework. The corresponding scalogram for the (d) noisy trace, (e) denoised trace obtained by the MRR method, and (f) denoised trace from the proposed framework.} 
\label{fig11} 
\end{figure}

The denoised data obtained by the MRR show a strong noise as indicated by arrows in Figures \ref{fig9}(b) and \ref{fig10}(b), while the denoised data corresponding to the proposed framework appear cleaner and smoother as shown in Figures \ref{fig9}(c) and \ref{fig10}(c). Additionally,  Figure \ref{fig11}(a) presents the 70$^{th}$ noisy channel, with corresponding denoised channels using the MRR and proposed framework as shown in Figures \ref{fig11}(b) and (c), respectively. Notably, the denoised channel obtained through the proposed framework exhibits lower noise levels compared to the MRR, with a clear P-wave arrival time.

For further analysis, Scalograms obtained from the continuous wavelet transform for the noisy and denoised channels are shown in Figures \ref{fig11}(d)-(f). Both Scalograms from the MRR and the proposed framework effectively preserve the frequency contents of the DAS signal. Although the MRR method effectively attenuates noise in SAFOD DAS data, Figure \ref{fig11}(e) indicates that some frequency content related to the noise still exists. This contrast is noticeable in the Scalogram of the proposed framework (Figure \ref{fig11}(f)), where these frequency contents are suppressed. 

\subsection{Fourth Field Example}
\begin{figure}[h] 
\centering
\includegraphics[scale=0.5]{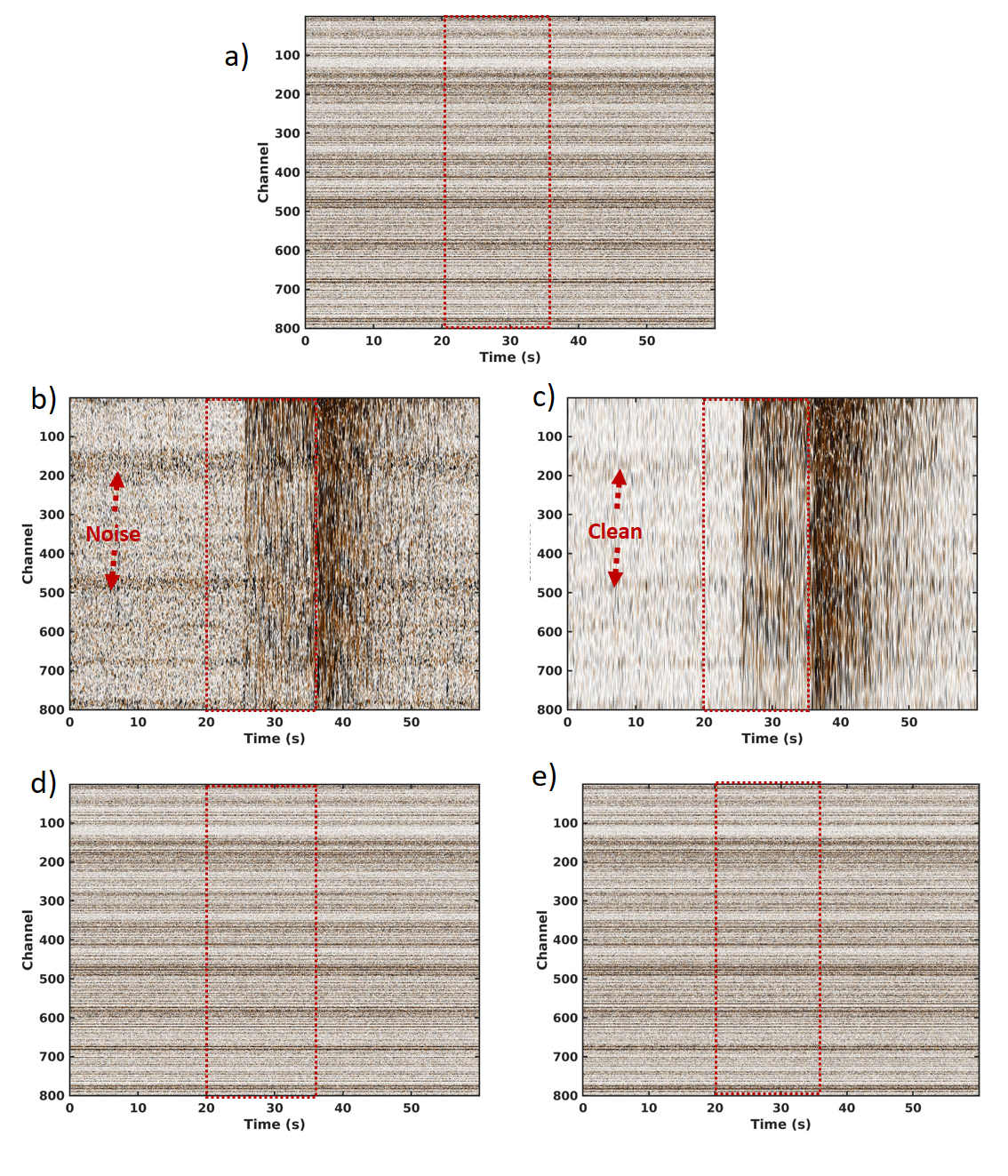} 
\caption{Denoising results for the fourth field example data from the SAFOD site. (a) Noisy DAS data. The denoised data correspond to the (b) MRR method and (c) the proposed framework. The removed noise sections correspond to the (d) MRR method and (e) the proposed framework.} 
\label{fig12} 
\end{figure}

\begin{figure}[h] 
\centering
\includegraphics[scale=0.5]{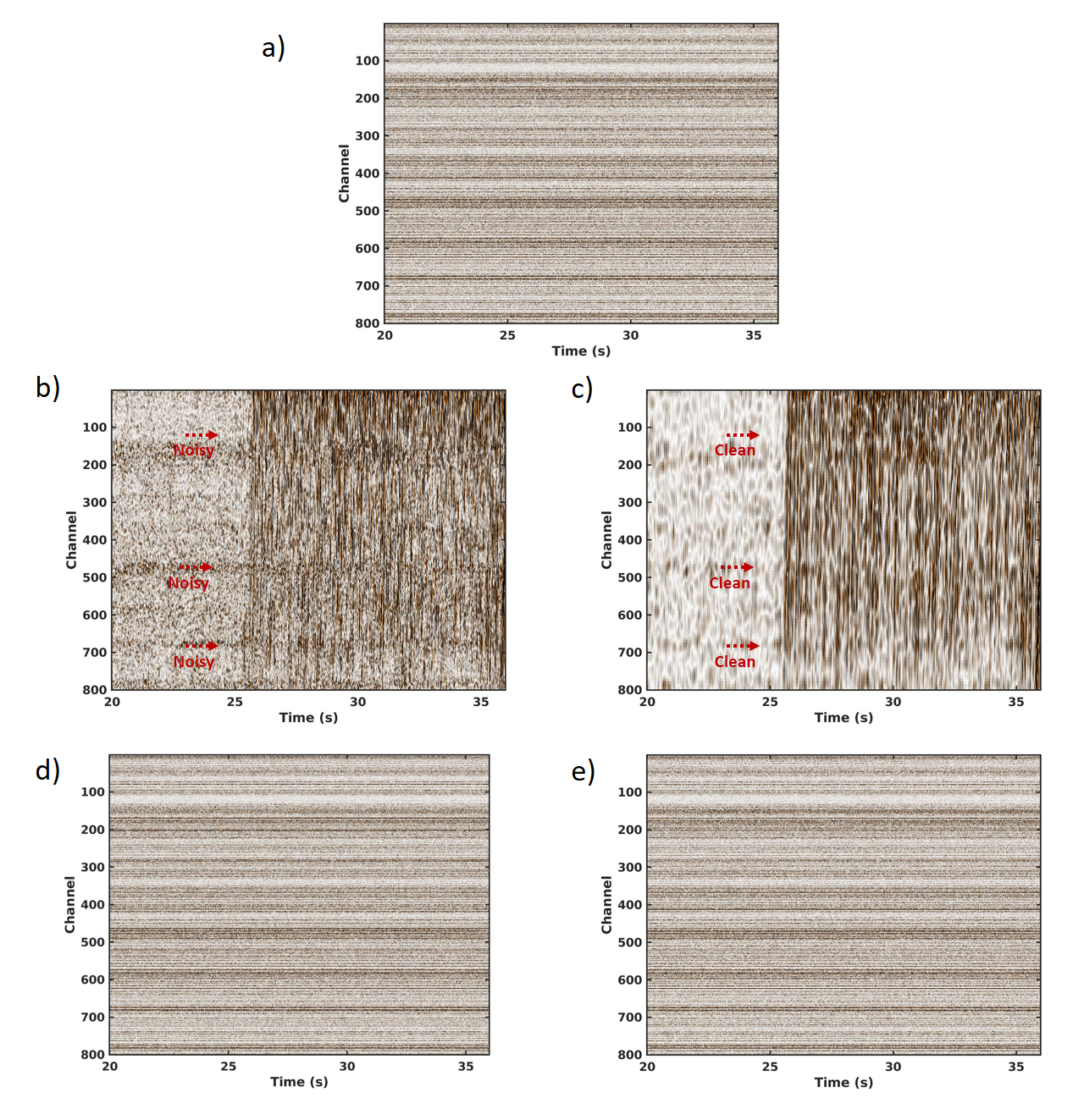} 
\caption{Zoomed sections from the red rectangles in Figure \ref{fig12}. (a) Noisy DAS data. The denoised data correspond to the (b) MRR method, and (c) the proposed framework. The removed noise sections correspond to the (d) MRR method and (e) the proposed framework.} 
\label{fig13} 
\end{figure}

Figure \ref{fig12}(a) shows the fourth field example extracted from SAFOD data, corresponding to an earthquake with a local magnitude of 2.86, denoted as "2017-07-02T03:17:32.300000Z\_mag2.86"  \citep{lellouch2021low}. Figures \ref{fig12}(b) and (c) show the denoised data obtained by the MRR and the proposed framework, respectively. The corresponding noise-removed sections for both methods are shown in Figures \ref{fig12}(d) and (e).  Zoomed-in sections for the noisy, denoised data and removed noise sections are shown in Figure \ref{fig13}. Notably, residual DAS noise persists in the denoised data obtained through the MRR method, as indicated by arrows in Figure \ref{fig13}(b). In contrast, the denoised data from the proposed framework appears cleaner as shown in Figure \ref{fig13}(c). Notably, the removed noise sections in Figure \ref{fig13} exhibit no signal leakages for both methods.

\begin{figure}[h] 
\centering
\includegraphics[scale=0.3]{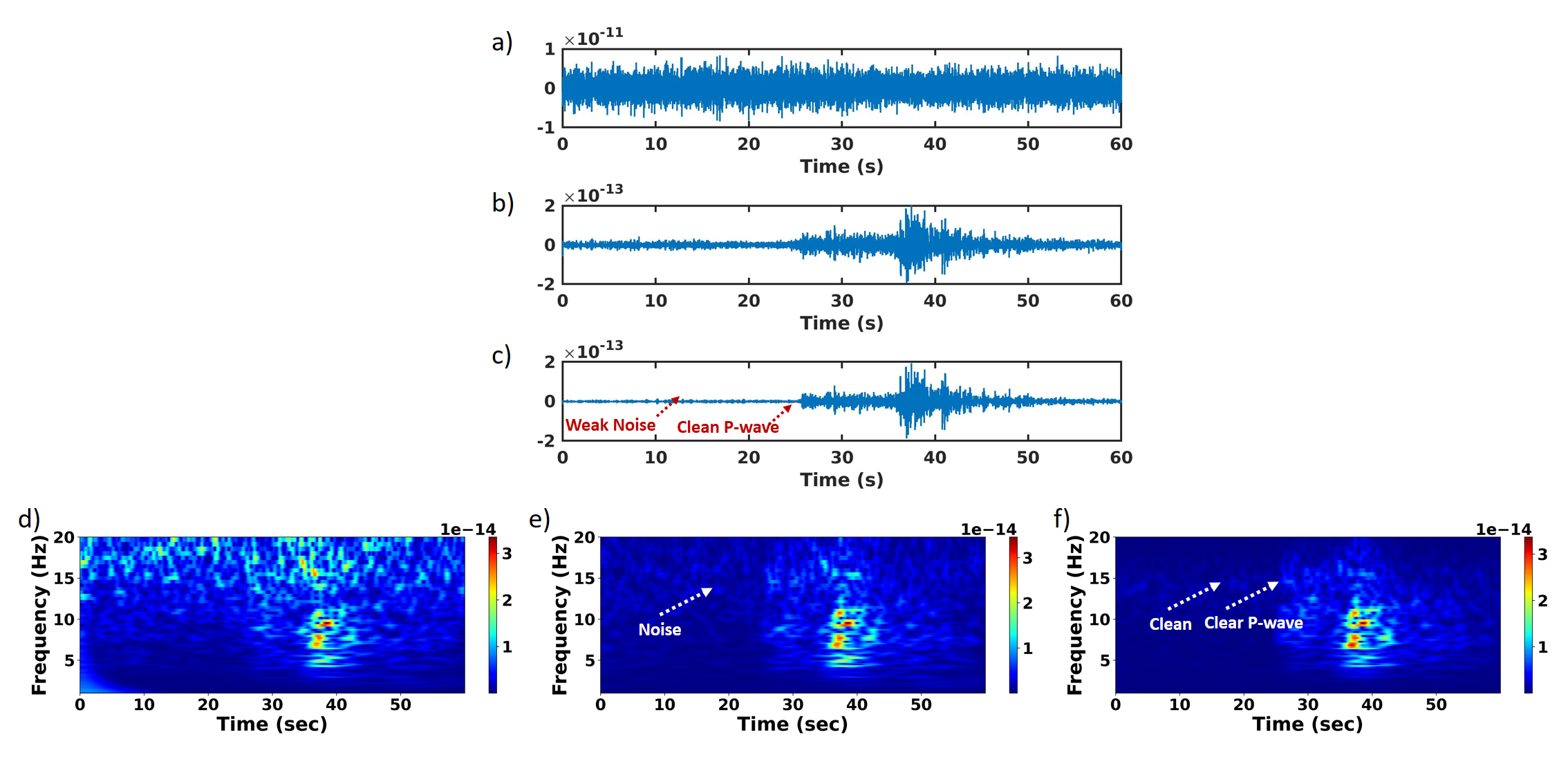} 
\caption{Single-trace compassion. (a) The 120 $^{th}$ noisy trace in Figure \ref{fig12}. The denoised traces correspond to (b) the MRR method and (c) the proposed framework. The corresponding scalogram for the (d) noisy trace, (e) denoised trace obtained by the MRR method, and (f) denoised trace from the proposed framework.} 
\label{fig14} 
\end{figure}

Figure \ref{fig14}(a) shows the 120$^{th}$ noisy channel, and the corresponding denoised channels using the MRR and the proposed framework are shown in Figures \ref{fig14}(b) and (c).  The denoised channel obtained from the proposed framework is cleaner, particularly preceding the P-wave arrival time, observable in both the Scalogram (Figure \ref{fig14}(f)) and the time domain representation (Figure \ref{fig14}(c)).  Both methods effectively recover the frequency contents associated with the DAS signal. However, the MRR method introduces some noise before the P-wave arrival time, as highlighted by arrows in Figure \ref{fig14}(e).

These results emphasize the capability of the proposed framework in attenuating complex DAS noise compared to the benchmark method, with minimal signal leakages.

\section{Discussion}
\subsection{Choosing The Optimal Network Architecture}
Given that the proposed deep learning model is designed to be plug-and-play, the choice of network architecture is crucial for robust denoising performance. A shallow architecture may struggle to reconstruct the DAS signal, while a deep architecture may entirely reconstruct the input data, encompassing the DAS noise. Consequently, we carefully select the optimal network architecture, employing the first field example for demonstration as follows:\\
1)  Initially, we focus on adjusting the patch size and the number of shift samples. Maintaining a fixed number of 4 encoder and decoder layers, we set the number of neurons in the first encoder layer to half of the patch size. For example, with a patch size of $16 \times 16$, the first encoder layer has 64 neurons (half of the product of the patch size). The number of neurons in the $i^{th}$ decoder layer is halved compared to the $(i-1)^{th}$ encoder layer. This strategy is maintained for the decoder, except the number of neurons is doubled instead of halved. We explore various patch sizes, including $16 \times 16$, $32 \times 32$, $48 \times 48$, and $64 \times 64$. Additionally, we test different shift samples, such as 1, 2, 4, 8, and 16. Accordingly, we find that using a patch size of $48 \time 48$ yields the lowest signal leakage and employing shift samples of 8 or fewer achieves optimal performance in attenuating DAS noise. To enhance model simplicity, we set the shift samples to 8.\\
2) Next, We try out different numbers of layers for the encoder and decoder (from 3 to 6) and test various numbers of neurons in each layer (128, 256, 512, 1024, and 2048). However,  adding more layers doesn't seem to make the denoising better. Also, using fewer neurons causes some signal leakage. Accordingly, we set the number of layers to three where the encoder layers have 1024, 512, and 256 neurons, while the decoder layers have 512, 1024, and 1024 neurons, respectively. Notable, using 2048 neurons does not enhance the results. \\
3) Finally, we fine-tune the number of fully connected layers for the compact layer. We experiment with 2, 3, 4, and 5 layers for each branch of the compact layer.  However, the denoising performance remains consistent across all trials, with no notable improvements in terms of signal leakage and noise attenuation. As a result, we establish the optimal architecture with two fully connected layers per branch.

\subsection{The Impact of Each Block in the Proposed Framework}
\begin{figure}[h] 
\centering
\includegraphics[scale=0.28]{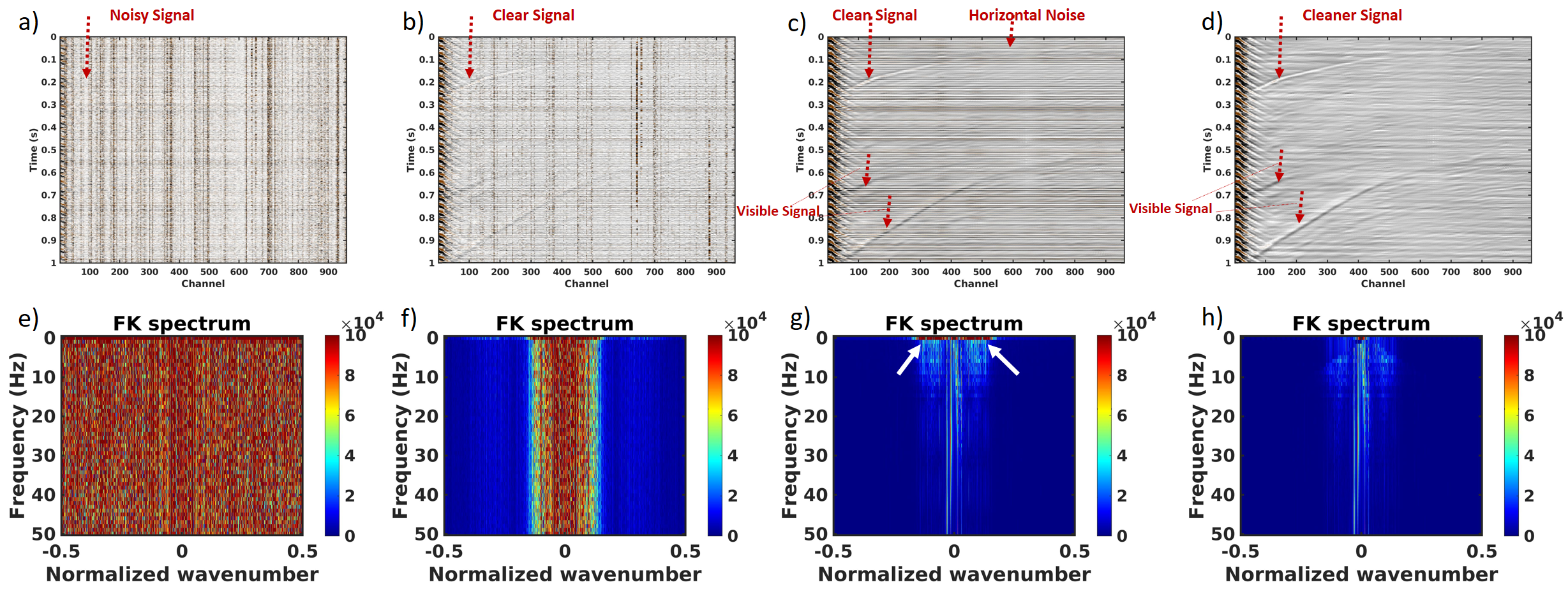} 
\caption{The output of each block in the proposed framework. (a) Noisy FORGE DAS data. (b) The data after applying the band-pass and median filters. (c) The DL output. (d) The output of the dip filter. The F-K spectra correspond to (e) the noisy data, (f) the data after applying the band-pass and median filters, (g) the DL output, and (h) the dip filter output.} 
\label{fig15} 
\end{figure}

To evaluate the impact of each block in the proposed framework, we apply the framework to FORGE DAS data named "FORGE\_78-32\_iDASv3P11\_UTC190426080738," associated with an earthquake with a local magnitude of -0.957 \citep{lellouch2021low}. The noisy data are shown in Figure \ref{fig15}(a), where the DAS signal is hardly visible. Subsequently, the band-pass filter removes high frequencies, rendering the DAS signal more visible (see Figure \ref{fig15}(b)). Afterward, the proposed deep learning model effectively attenuates the majority of DAS noise, resulting in a cleaner signal (see Figure \ref{fig15}(c)). However, some residual horizontal noise persists, highlighted by arrows in the corresponding F-K spectrum (see Figure \ref{fig15}(g)). Finally, the dip filter suppresses the remaining horizontal noise as shown in Figure \ref{fig15}(d). The DAS signal becomes distinctly visible and cleaner, as observed in the frequency content of the F-K spectrum shown in Figures \ref{fig15}(g) and (f).

\subsection{The Effect of the Training Ratio}
\begin{figure}[h] 
\centering
\includegraphics[scale=0.3]{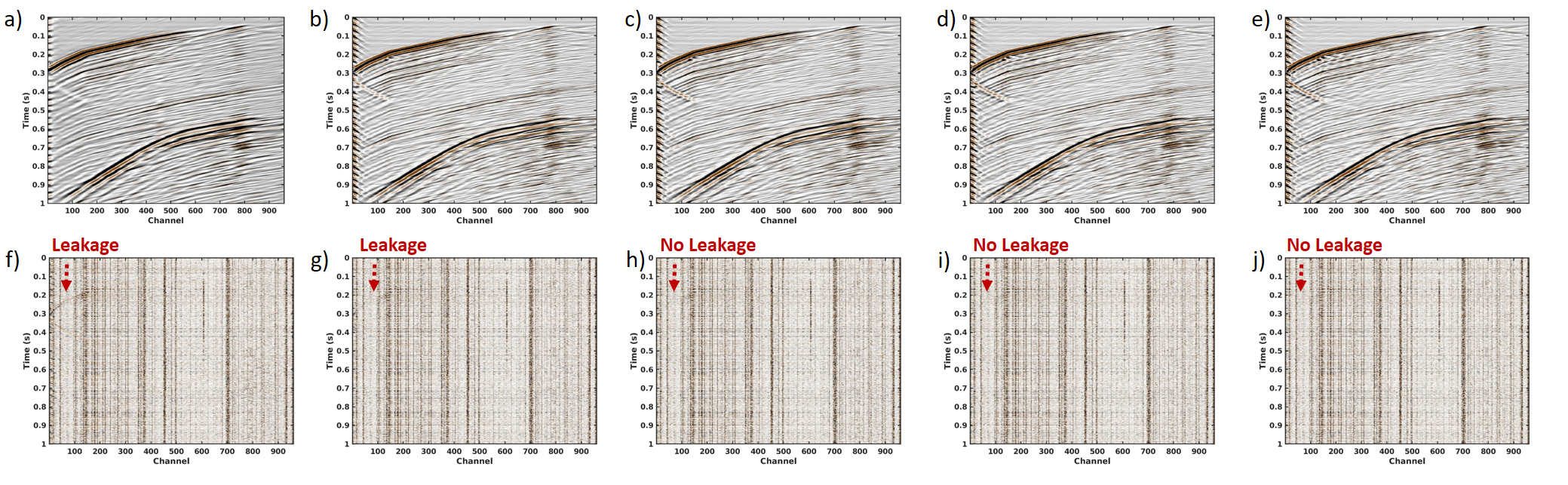} 
\caption{The impact of the training ratio. The denoised data using a training ratio of (a) 10\%, (b) 30\%, (c) 50\%, (d) 70\%, and (e) 90\%. (f-j) The corresponding removed noise sections.} 
\label{fig16} 
\end{figure}

As the proposed DL network works in an unsupervised scheme, it dynamically optimizes its parameters for each new dataset. In this section, we assess the performance of the proposed framework using varying training ratios where the first field data are used in this evaluation. Specifically, we test the denoising performance of the proposed algorithm using training ratios of 10\%, 30\%, 50\%, 70\%, and 90\%. Figure \ref{fig16} shows the denoised and removed noise sections for each case.

Notably, when the training ratio is set at 30\% or lower, signal leakage appears in the corresponding removed noise section, highlighted by arrows in Figures \ref{fig16}(f) and (g). However, when the training ratio is increased to 50\% or higher, no signal leakages are observed, as shown in Figures \ref{fig16}(h)-(i).

\subsection{Computational Complexity}
The proposed DL network required 456 seconds to denoise the first field example using all the patches. In comparison, the FCDensenet \citep{yang2023denoising} took 1300 seconds to denoise the same field data. 

Moreover, by decreasing the number of patches involved in the optimization process, there was a notable reduction in the computational complexity of the proposed DL model. For the first field data, the DL model requires 425 seconds to optimize its network parameters using 90\% of the extracted patches. However, the consumed time is significantly reduced to 242 seconds when only 50\% of the patches are utilized. This strategy proves effective in reducing the computational complexity of the proposed DL network while maintaining the same denoising performance. These results were obtained using a single GPU Nvidia GeForce RTX 2080 Ti.

\subsection{Interpretation of the Deep Learning Model}
In this section, we interpret the performance of the proposed DL model by analyzing the weight matrices inherent in the proposed architecture. Given the proposed DL model comprising multiple fully connected layers, the weight matrices governing the connections between each adjacent layer are considered as generalized features extracted by the model. For example, the input flattened patches, with a size of 2304, are connected to the first encoder layer comprising 1024 neurons, resulting in weight matrices of dimensions $2304 \times 1024$. To facilitate visualization, these weight matrices are reshaped into a 2D format, resulting in matrices of dimensions $48 \times 48 \times 1024$. This reshaping yields 1024 features, each maintaining the same size as the input data.

\begin{figure}[h] 
\centering
\includegraphics[scale=0.3]{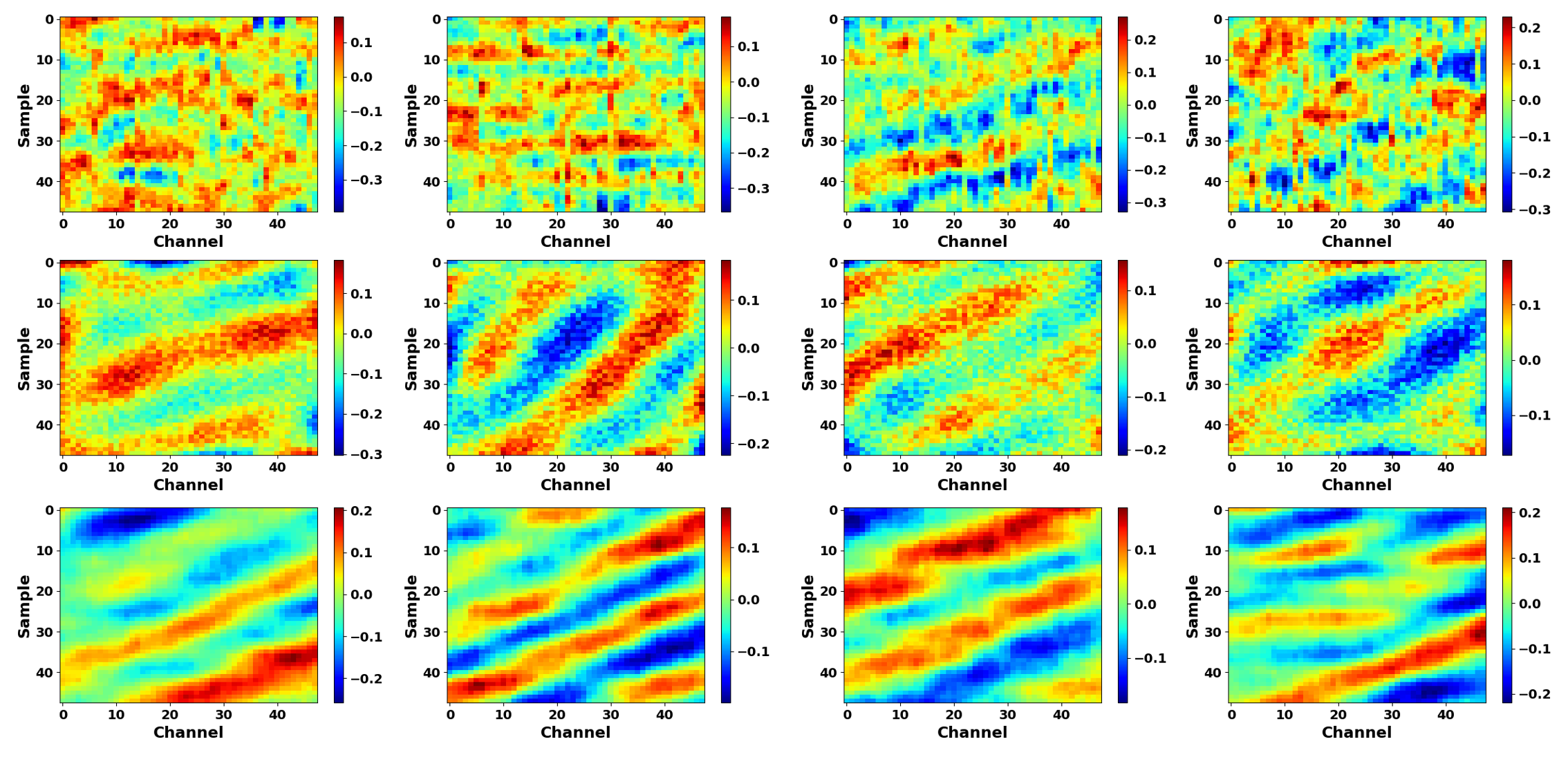} 
\caption{The weight matrices associated with the deep learning model.  In the first row, a subset of weight matrices for the connections between the input layer and the first encoder layer. The second row displays a subset of weights matrices between the finest scale and the first encoder layer. The third row signifies a subset of weight matrices representing the connections between the last decoder layer and the output layer. Note that the samples in each row are randomly selected.} 
\label{fig17} 
\end{figure}

The first row of Figure \ref{fig17} denotes some of the weight matrices associated with the band-pass filtered data, while the second row of Figure \ref{fig17} shows the weight matrices delineating the connections between the scale patches and the first encoder layer. It is noticeable that the weight matrices, corresponding to the band-pass filtered data, exhibit strong DAS noise, with the DAS signal being hardly visible. In contrast, the weight matrices corresponding to the scale input reveal a visible representation of the DAS signal.

Subsequently, the network undergoes a learning process aimed at extracting the DAS signal from the band-pass filtered data, guided by the data obtained at the scale level. The third row of Figure \ref{fig17} shows a subset of the weight matrices between the last decoder layer and the output layer, wherein the signal attains a heightened level of clarity and smoothness in comparison to the weight matrices at the input side.

This analysis reveals that the proposed deep learning model tunes its parameters to reconstruct the DAS signal, as apparent in the weight matrices associated with the input scale, while concurrently attenuating the DAS noise that exists in the weight matrices associated with the band-pass filtered data. Note that the samples in each row of Figure \ref{fig17} are chosen randomly.

\subsection{The Limitations of the Proposed Framework}
\begin{figure}[h] 
\centering
\includegraphics[scale=0.35]{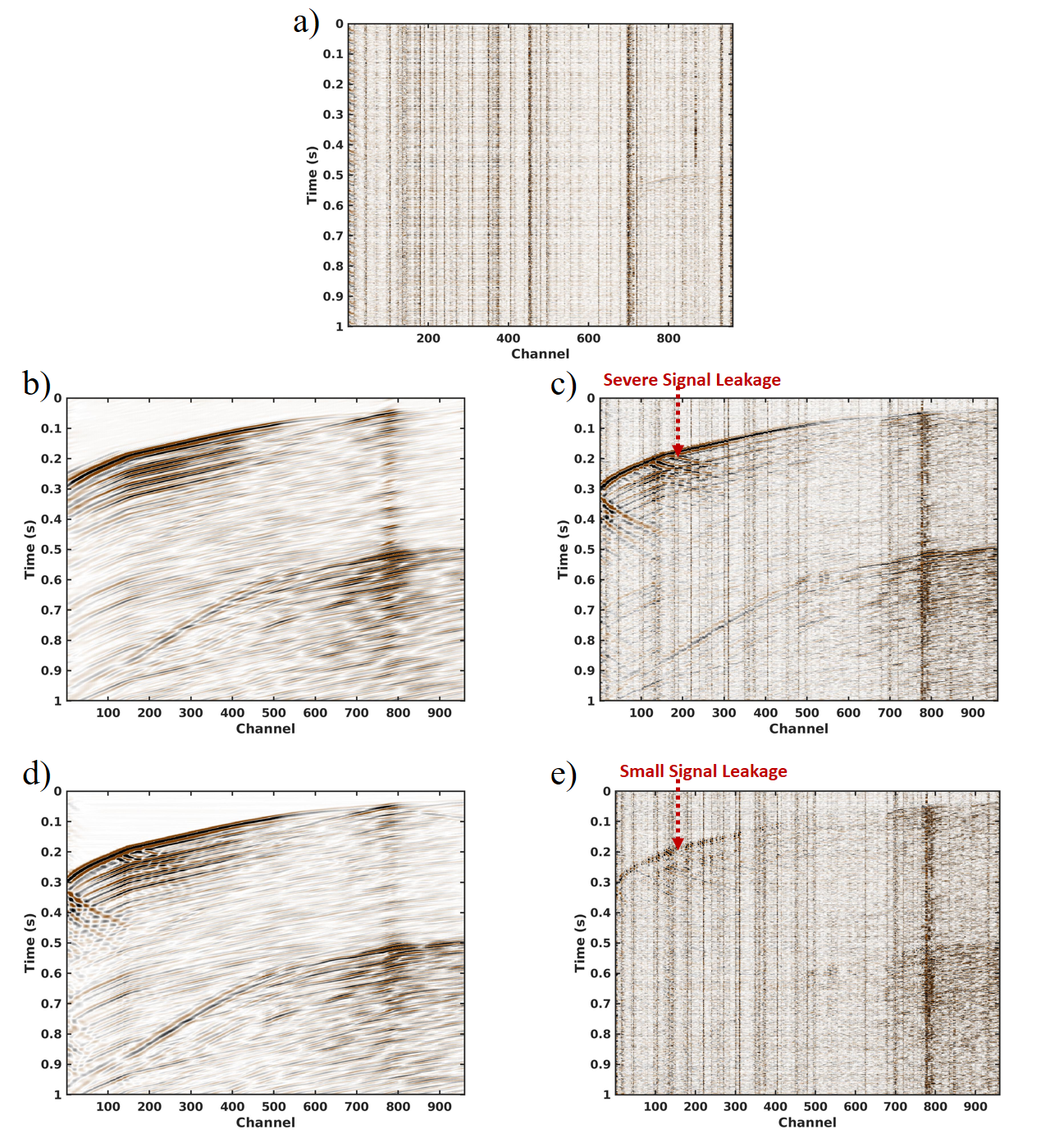} 
\caption{An example of signal leakage from the FORGE data. (a) Noisy DAS data. (b) The denoised data and (c) the removed noise section obtained by the FCDensenet model. (d) The denoised data and (e) the removed noise section corresponding to the proposed framework.} 
\label{fig18} 
\end{figure}

To further investigate the denoising performance of the proposed framework, we analyze additional FORGE data, named "FORGE\_78-32\_iDASv3-P11\_UTC190423213209.sgy," corresponding to an earthquake with a local magnitude of 0.91, as shown in Figure \ref{fig18}(a) \citep{lellouch2021low}. Denoised data from both the FCDensenet model and the proposed framework are shown in Figures \ref{fig18}(b) and \ref{fig18}(d), respectively. Figures \ref{fig18}(c) and \ref{fig18}(e) show the removed noise sections by both methods. Both methods effectively attenuate the DAS noise, but there is some signal leakage as indicated by arrows in Figures \ref{fig18}(c) and \ref{fig18}(e). The proposed framework has less signal leakage compared to the FCDensenet model, where the signal leakage corresponding to the FCDensenet model is severe. The signal leakage observed is likely a result of amplitude variations between the obtained scale and the band-pass filtered data. This variation poses a challenge for guiding the DL model in accurately reconstructing the DAS signals.

\begin{figure}[h] 
\centering
\includegraphics[scale=0.25]{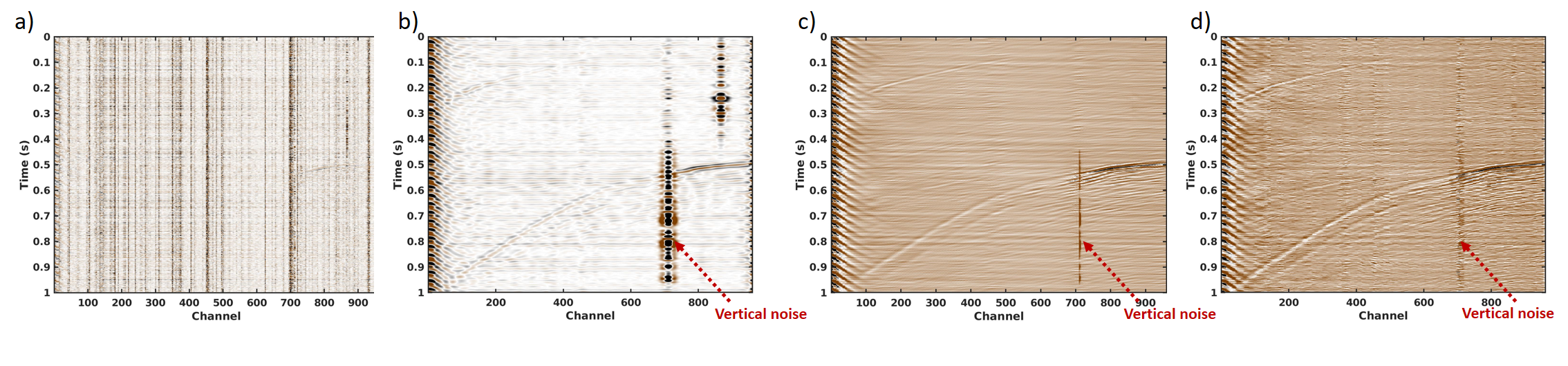} 
\caption{An example of remaining vertical noise from the FORGE data. (a) Noisy DAS data. (b) The corresponding finest scale. The denoised data correspond to (c) the proposed framework and (d) the SOMF method.} 
\label{fig19} 
\end{figure}

In addition, we assess the efficacy of the proposed framework using another data from FORGE, named "FORGE\_78-32\_iDASv3-P11\_UTC190428180038.sgy," associated with an earthquake having a local magnitude of 1.07 \citep{lellouch2021low}. Figures \ref{fig19}(a)-(d) show the noisy data, the corresponding finest scale, the denoised data from the proposed framework, and the denoised data from the SOMF method, respectively. The proposed framework effectively attenuates the noise in the DAS data compared to the SOMF method. However, there remains residual vertical noise in the removed noise section, indicated by an arrow in Figure \ref{fig19}(c). This is likely due to the presence of strong vertical noise in the finest scale guiding the DL network, leading the network to reconstruct this erratic noise as a signal. Although the log cosh loss function diminishes the amplitude of this erratic noise, the DL model still reconstructs a portion of it.

The provided examples highlight the significant influence of the finest scale in attenuating DAS noise. To improve the denoising performance, we can guide the network with better-quality data, aiming to minimize signal leakages and better preserve the DAS signal.

\section{Conclusions}
Distributed Acoustic Sensing (DAS) introduces a new technology for obtaining seismic data with enhanced resolution. This innovation offers a more detailed comprehension of the subsurface structures. DAS data are characterized by a low signal-to-noise (SNR) ratio due to the complex noise associated with fiber-optic cables. In this study, we proposed an unsupervised deep learning model guided by a time-frequency representation of the data to extract the DAS signal and iteratively attenuate DAS noise. We employ a patching technique to extract patches from the band-pass filtered data and the smallest scale of the continuous wavelet transform (CWT). A self-attention network is proposed to capture dependencies between the band-pass filtered data and the CWT scale. Additionally, we use the Log cosh as a loss function to compare reconstructed data with the band-pass filtered data, serving as the target for the network. Following this, we apply a dip filter in the f-k domain to suppress any remaining horizontal noise. 

Testing the proposed framework on field examples from San Andreas Fault Observatory at Depth (SAFOD) and Frontier Observatory for Research in Geothermal Energy (FORGE) datasets demonstrates robust denoising performance, with minimal signal leakages and cleaner denoised data compared to benchmark methods. The proposed framework exhibits a robust generalization across diverse datasets, where the DL network tunes its parameters for each new dataset without labels. Furthermore, our analysis indicates the efficiency of the DL network in learning with a small number of patches, leading to a reduction in computational complexity. 
In addition, our analysis emphasizes the importance of improving the quality of guiding data to enhance the denoising performance of the DL network.

\section{Acknowledgment}
This publication is based on work supported by the King Abdullah University of Science and Technology (KAUST). The authors thank the DeepWave sponsors for their support.

\bibliographystyle{unsrtnat}
\bibliography{references}  %

\newpage
\listoffigures

\newpage

\newpage

\newpage

\newpage

\newpage

\newpage

\newpage

\newpage

\newpage

\newpage

\newpage

\newpage

\newpage

\newpage

\newpage

\newpage

\newpage

\newpage

\newpage

\end{document}